\documentclass[aps,showpacs]{revtex4}
\usepackage{graphicx}
\begin{document}

\title{Transverse energy per charged particle at relativistic
energies from a statistical model with expansion}
\author{Dariusz Prorok}
\affiliation{Institute of Theoretical Physics, University of
Wroc{\l}aw,\\ Pl.Maksa Borna 9, 50-204  Wroc{\l}aw, Poland}
\date{September 15, 2004}

\begin{abstract}
Transverse energy and charged particle pseudorapidity densities at
midrapidity and their ratio, $dE_{T}/d\eta\mid_{mid} /
dN_{ch}/d\eta\vert_{mid}$, are evaluated in a statistical model
with longitudinal and transverse flows for the wide range of
colliders, from AGS to RHIC at $\sqrt{s_{NN}}=200$ GeV.
Evaluations are done at freeze-out parameters established
independently from fits to observed particle yields and $p_{T}$
spectra. Decays of hadron resonances are treated thoroughly and
are included in derivations of $dE_{T}/d\eta\vert_{mid}$ and
$dN_{ch}/d\eta\vert_{mid}$. The predictions of the model agree
well with the experimental data. However, some (explicable)
overestimation of the ratio has been observed.
\end{abstract}

\pacs{25.75.-q, 25.75.Dw, 24.10.Pa, 24.10.Jv} \maketitle

\section {Introduction }

In this paper, the idea of an independent test of the
applicability of a statistical model for the description of the
soft part of particle production in a heavy-ion collision
postulated in \cite{Prorok:2002ta}, is developed for the much more
realistic case of a hadron gas and its expansion. So far, the
statistical model has been applied successfully in description of
particle yield ratios and $p_{T}$ spectra measured in heavy-ion
collisions
\cite{Braun-Munzinger:1994xr,Braun-Munzinger:1995bp,Cleymans:1996cd,Stachel:wh,Braun-Munzinger:1999qy,Becattini:2000jw,Braun-Munzinger:2001ip,Florkowski:2001fp,Broniowski:2001we,Broniowski:2001uk,Baran:2003nm,Broniowski:2002nf,Michalec:2001um,Broniowski:2002am}
(there are also computation packages for thermal studies available
from the Web \cite{Torrieri:2004zz,Wheaton:2004qb}). Now, the
freeze-out parameters obtained from those analyses will be used to
evaluate global observables: the transverse energy density
$dE_{T}/d\eta$, the charged particle multiplicity density
$dN_{ch}/d\eta$ and their ratio. The advantage of such an approach
is based on the fact that transverse energy measurements are
independent of hadron spectroscopy (in particular, no particle
identification is necessary), therefore they could be used as an
additional test of the self-consistency of a statistical model.
The same holds true for the charged particle multiplicity, which
actually is the charged \emph{hadron} multiplicity, according to
the experimental definition given in \cite{Back:2000gw}.

The experimentally measured transverse energy is defined as

\begin{equation}
E_{T} = \sum_{i = 1}^{L} \hat{E}_{i} \cdot \sin{\theta_{i}} \;,
\label{Etdef}
\end{equation}
%%%%%%%%%%%%%%%%%%%%%%%%%%%%eq.1

\noindent where $\theta_{i}$ is the polar angle, $\hat{E}_{i}$
denotes $E_{i}-m_{N}$ ($m_{N}$ means the nucleon mass) for baryons
and the total energy $E_{i}$ for all other particles, and the sum
is taken over all $L$ emitted particles \cite{Adcox:2001ry}.
Additionally, in the case of RHIC at $\sqrt{s_{NN}}=200$ GeV,
$E_{i}+m_{N}$ is taken instead of $E_{i}$ for antibaryons
\cite{Bazilevsky:2002fz}.

The statistical model with single freeze-out is used (for details
see \cite{Broniowski:2002nf} and references therein). The model
reproduces very well ratios and $p_{T}$ spectra of particles
measured at RHIC
\cite{Florkowski:2001fp,Broniowski:2001we,Broniowski:2001uk}. The
main assumption of the model is the simultaneous occurrence of
chemical and thermal freeze-outs, which is important if $p_{T}$
spectra are considered (this enables to neglect the possible
elastic interactions after the chemical freeze-out). Since in the
present paper the integrated quantities over $p_{T}$ are dealt
with, the above-mentioned assumption should not be so important
for final results.

The actually detected (stable) particles have two sources:
(\textit{a}) a thermal gas and (\textit{b}) secondaries produced
by decays and sequential decays of primordial resonances. All
stable hadrons and confirmed resonances up to a mass of $2$ GeV
from the Particle Data Tables \cite{Hagiwara:fs} are constituents
of the gas. The distributions of particles from source
(\textit{a}) are given by a Bose-Einstein or a Fermi-Dirac
distribution at the freeze-out. The distributions of secondaries
(source (\textit{b})) can be obtained from the elementary
kinematics of a many-body decay or from the superposition of two
or more such decays (for details see the Appendix and
\cite{Broniowski:2002nf}). In the following, all possible (2-, 3-
and 4-body) decays with branching ratios not less than $1 \%$ are
considered. Also almost all possible sequential decays are taken
into account, namely: $2\circ2$, $2\circ2\circ2$,
$2\circ2\circ2\circ2$, $2\circ3$, $2\circ4$, $3\circ2$, $3\circ3$,
$2\circ2\circ3$, $2\circ3\circ2$, $3\circ2\circ2$,
$2\circ3\circ3$, where 2, 3 and 4 mean the 2-, 3- and 4-body decay
respectively, and a cascade proceeds from the right to the left
(as in the usual mathematical definition of the superposition of
functions). It should be stressed that all contributions from weak
decays are included. The contribution to the transverse energy
from the omitted cascades has been estimated at $0.2 \%$. But
since the most of these cascades ends with two photons, they do
not contribute to the charged particle multiplicity at all.

\section { The basis of the freeze-out model }
\label{Foundat}

The following are the foundations of the model. A noninteracting
gas of stable hadrons and resonances at chemical and thermal
equilibrium is created at the Central Rapidity Region (CRR) of a
collision. The gas cools and expands, and after reaching the
freeze-out point it ceases. The conditions for the freeze-out are
expressed by values of two independent thermal parameters: the
temperature $T$ and the baryon number chemical potential
$\mu_{B}$. The strangeness chemical potential $\mu_{S}$ is
determined from the requirement that the overall strangeness of
the gas equals zero.

A freeze-out hypersurface is defined by the condition

\begin{equation}
\tau = \sqrt{t^{2}-r_{x}^{2}-r_{y}^{2}-r_{z}^{2}}= const \;,
\label{Hypsur}
\end{equation}
%%%%%%%%%%%%%%%%%%%%%%%%%%%%eq.2

\noindent which means that the freeze-out takes place at a fixed
moment of the invariant time $\tau$. Additionally, it is assumed
that the four-velocity of an element of the freeze-out
hypersurface is proportional to its coordinate,

\begin{equation}
u^{\mu}={ {x^{\mu}} \over \tau}= {t \over \tau}\; \left(1,{
{r_{x}} \over t},{{r_{y}} \over t},{{r_{z}} \over t}\right) \;.
\label{Velochyp}
\end{equation}
%%%%%%%%%%%%%%%%%%%%%%%%%%%%eq.3

\noindent Then the following parameterization of the hypersurface
is chosen:

\begin{equation}
t= \tau \cosh{\alpha_{\parallel}}\cosh{\alpha_{\perp}},\;\;\;
r_{x}=  \tau \sinh{\alpha_{\perp}}\cos{\phi},\;\;\; r_{y}=  \tau
\sinh{\alpha_{\perp}}\sin{\phi},\;\;\;r_{z}=\tau
\sinh{\alpha_{\parallel}}\cosh{\alpha_{\perp}}, \label{Parahyp}
\end{equation}
%%%%%%%%%%%%%%%%%%%%%%%%%%%%eq.4

\noindent where $\alpha_{\parallel}$ is the rapidity of the
element, $\alpha_{\parallel}= \tanh^{-1}(r_{z}/t)$, and
$\alpha_{\perp}$ determines the transverse radius

\begin{equation}
r= \sqrt{r_{x}^{2}+r_{y}^{2}}= \tau \sinh{\alpha_{\perp}}.
\label{Transsiz}
\end{equation}
%%%%%%%%%%%%%%%%%%%%%%%%%%%%eq.5

\noindent To keep the transverse size finite, $r$ is restricted by
the condition $r < \rho_{max}$. In this way one has two additional
parameters of the model, $\tau$ and $\rho_{max}$, connected with
the geometry of the freeze-out hypersurface.

Also the transverse velocity, $v_{\rho}$, can be obtained

\begin{equation}
v_{\rho}= \sqrt{\left({{r_{x}} \over t}\right)^{2}+\left({{r_{y}}
\over t}\right)^{2}}={ {\tau \sinh{\alpha_{\perp}}} \over t} = {
{\tanh{\alpha_{\perp}}} \over {\cosh{\alpha_{\parallel}}} } = {
{\beta_{\perp}} \over {\cosh{\alpha_{\parallel}}} }\; ,
\label{Transvel}
\end{equation}
%%%%%%%%%%%%%%%%%%%%%%%%%%%%eq.6

\noindent which is the value of the transverse velocity
$\beta_{\perp}$ from the central slice after boosting it in the
longitudinal direction. The transverse velocity can be expressed
as a function of the transverse radius

\begin{equation}
\beta_{\perp}(r)= \tanh{\alpha_{\perp}}= { r \over
{\sqrt{\tau^{2}+r^{2}}}}\;. \label{Betprof1}
\end{equation}
%%%%%%%%%%%%%%%%%%%%%%%%%%%%eq.7

\noindent Since it is an increasing function of $r$, the maximum
value of $\beta_{\perp}$ called the maximum transverse-flow
parameter (or the surface velocity), is given by

\begin{equation}
\beta_{\perp}^{max}= { \rho_{max} \over
{\sqrt{\tau^{2}+\rho_{max}^{2}}}}= { {\rho_{max}/\tau} \over
{\sqrt{1+(\rho_{max}/\tau)^{2}}}}\;, \label{Betmax}
\end{equation}
%%%%%%%%%%%%%%%%%%%%%%%%%%%%eq.8

\noindent so it depends only on the ratio $\rho_{max}/\tau$.

\section {Transverse energy and charged particle densities}
\label{density}

According to the general description founded in
\cite{Cooper:1974mv} and developed in
\cite{Broniowski:2001we,Broniowski:2001uk} for the case with
decays taken into account, the invariant distribution of the
measured particles of species $i$ has the form

\begin{equation}
{ {dN_{i}} \over {d^{2}p_{T}\;dy} }=\int
p^{\mu}d\sigma_{\mu}\;f_{i}(p \cdot u) \;, \label{Cooper}
\end{equation}
%%%%%%%%%%%%%%%%%%%%%%%%%%%%eq.9

\noindent where $d\sigma_{\mu}$ is the normal vector on a
freeze-out hypersurface, $p \cdot u = p^{\mu}u_{\mu}$ , $u_{\mu}$
is the four-velocity of a fluid element and $f_{i}$ is the final
momentum distribution of the particle in question. The final
distribution means here that $f_{i}$ is the sum of primordial and
simple and sequential decay contributions to the particle
distribution (for details see \cite{Broniowski:2002nf}). For the
hypersurface and expansion described in sect.~\ref{Foundat},
eq.~(\ref{Cooper}) takes the following form:

\begin{equation}
{ {dN_{i}} \over {d^{2}p_{T}\;dy} }= \tau^{3}\;
\int\limits_{-\infty}^{+\infty}
d\alpha_{\parallel}\;\int\limits_{0}^{\rho_{max}/\tau}\;\sinh{\alpha_{\perp}}
d(\sinh{\alpha_{\perp}})\; \int\limits_{0}^{2\pi} d\xi\;p \cdot u
\; f_{i}(p \cdot u) \;, \label{Cooper2}
\end{equation}
%%%%%%%%%%%%%%%%%%%%%%%%%%%%eq.10

\noindent where

\begin{equation}
p \cdot u = m_{T}\cosh{\alpha_{\parallel}}\cosh{\alpha_{\perp}}-
p_{T}\cos{\xi}\sinh{\alpha_{\perp}}\;. \label{Peu}
\end{equation}
%%%%%%%%%%%%%%%%%%%%%%%%%%%%eq.11

\noindent Note that the distribution expressed by
eqs.~(\ref{Cooper2}) and (\ref{Peu}) is explicitly boost invariant
(in fact, it is constant with respect to rapidity).

The rapidity density of particle species $i$ is given by

\begin{equation}
{ {dN_{i}} \over {dy} } = \int d^{2}p_{T}\; { {dN_{i}} \over
{d^{2}p_{T}\;dy} }\;, \label{Parapdens}
\end{equation}
%%%%%%%%%%%%%%%%%%%%%%%%%%%%eq.12

\noindent whereas the corresponding pseudorapidity density reads

\begin{equation}
{ {dN_{i}} \over {d\eta} } = \int d^{2}p_{T}\; {{dy} \over {d\eta}
} \; { {dN_{i}} \over {d^{2}p_{T}\;dy} }= \int d^{2}p_{T}\; {p
\over {E_{i}} } \; { {dN_{i}} \over {d^{2}p_{T}\;dy} }\;.
\label{Partdens}
\end{equation}
%%%%%%%%%%%%%%%%%%%%%%%%%%%%eq.13

\noindent Analogously, the transverse energy pseudorapidity
density for the same species can be written as

\begin{equation}
{ {dE_{T,i}} \over {d\eta} } = \int d^{2}p_{T}\; \hat{E}_{i} \cdot
{{p_{T}} \over p} \; {{dy} \over {d\eta} }\; { {dN_{i}} \over
{d^{2}p_{T}\;dy} }= \int d^{2}p_{T}\;{p_{T}} \; { {\hat{E}_{i}}
\over {E_{i}} }\; { {dN_{i}} \over {d^{2}p_{T}\;dy} }\;.
\label{Etraden}
\end{equation}
%%%%%%%%%%%%%%%%%%%%%%%%%%%%eq.14

\noindent For the quantities at midrapidity one has

\begin{equation}
{ {dN_{i}} \over {d\eta} }\;\Big\vert_{mid}= \int d^{2}p_{T}\;{
{dN_{i}} \over {d^{2}p_{T}\;dy} }\; {
{\sqrt{p_{T}^{2}+v_{c.m.s}^{2}m_{i}^{2}}} \over {m_{T}} } \;,
\label{Partdenmid}
\end{equation}
%%%%%%%%%%%%%%%%%%%%%%%%%%%%eq.15

\begin{equation}
{ {dE_{T,i}} \over {d\eta} }\;\Big\vert_{mid} = \cases{ \int
d^{2}p_{T}\;{p_{T}} \;{ {dN_{i}} \over {d^{2}p_{T}\;dy} }\;  {
{m_{T}-\sqrt{1-v_{c.m.s}^{2}}m_{N}} \over {m_{T}} }, i=nucleon
 \cr \cr \int d^{2}p_{T}\;{p_{T}} \;{ {dN_{i}} \over
{d^{2}p_{T}\;dy} }, i \neq nucleon\;.} \label{Etdenmid}
\end{equation}
%%%%%%%%%%%%%%%%%%%%%%%%%%%%eq.16

\noindent where $v_{c.m.s}$ is the velocity of the center of mass
of two colliding nuclei with respect to the laboratory frame (only
for RHIC $v_{c.m.s}=0$). For RHIC at $\sqrt{s_{NN}}=200$ GeV the
case $i \neq nucleon$ in eq.~(\ref{Etdenmid}) is replaced by

\begin{equation}
{ {dE_{T,i}} \over {d\eta} }\;\Big\vert_{mid} = \cases{ \int
d^{2}p_{T}\;{p_{T}} \;{ {dN_{i}} \over {d^{2}p_{T}\;dy} }\;  {
{m_{T}+m_{N}} \over {m_{T}} }, i=antinucleon
 \cr \cr \int d^{2}p_{T}\;{p_{T}} \;{ {dN_{i}} \over
{d^{2}p_{T}\;dy} }, i \neq nucleon,\;antinucleon\;.}
\label{Etdenmidb}
\end{equation}
%%%%%%%%%%%%%%%%%%%%%%%%%%%%eq.17

Now, the overall charged particle and transverse energy densities
can be obtained

\begin{equation}
{ {dN_{ch}} \over {d\eta} }\;\Big\vert_{mid}= \sum_{i \in B} {
{dN_{i}} \over {d\eta} }\;\Big\vert_{mid}\;, \label{Nchall}
\end{equation}
%%%%%%%%%%%%%%%%%%%%%%%%%%%%eq.18

\begin{equation}
{ {dE_{T}} \over {d\eta} }\;\Big\vert_{mid}= \sum_{i \in A} {
{dE_{T,i}} \over {d\eta} }\;\Big\vert_{mid} \;, \label{Etall}
\end{equation}
%%%%%%%%%%%%%%%%%%%%%%%%%%%%eq.19

\noindent where $A$ and $B$ ($B \subset A$) denote sets of species
of finally detected particles. In the view of the definition given
in \cite{Back:2000gw} and the detailed description of the
experimental setup and the analysis procedure from
\cite{Adcox:2000sp}, the set of charged particles $B$ can consist
of stable hadrons only, $B=\{\pi^{+},\; \pi^{-},\; K^{+},\;
K^{-},\; p,\; \bar{p}\}$, whereas $A$ also includes photons,
$K_{L}^{0},\; n$ and $\bar{n}\;$ \cite{Adcox:2001ry}.

\section {Results}
\label{Finl}

To check the self-consistency of the described model, rapidity
densities of pions, kaons, protons and antiprotons have been
calculated for the $5 \%$ most central Au-Au collisions at
$\sqrt{s_{NN}}=130$ GeV at RHIC. The supplied values of the
thermal and geometric parameters are in this case $T = 165$ MeV,
$\mu_{B} = 41$ MeV, $\tau = 8.2$ fm and $\rho_{max} = 6.9$ fm
\cite{Florkowski:2001fp,Broniowski:2002nf}. The geometric
parameters were obtained from the fit to the $p_{T}$ spectra of
the above-mentioned particles \cite{Adcox:2001mf}. The integrated
yields over $p_{T}$ are also given in \cite{Adcox:2001mf}, so the
comparison with the predictions of eq.~(\ref{Parapdens}) can be
done easily. The results are presented in TABLE\,\ref{Table1}.
Note that the very good agreement has been found.

The presentation of the main results of the paper needs a few
comments concerning the AGS case. In the all cited papers the same
method of establishing the thermal parameters $T$ and $\mu_{B}$ is
applied. The method is based on the best fit of calculated
particle density ratios to the relative particle abundance data.
But the different model of the freeze-out was applied for the
description of $p_{T}$ spectra measured at AGS
\cite{Braun-Munzinger:1994xr,Stachel:wh}. In that model (for
details see \cite{Schnedermann:1993ws}), the freeze-out happens
instantaneously in the $r$ direction, \emph{i.e.} at a constant
value of $t$ (\emph{not} at a constant value of $\tau$ as here).
The shape of a hypersurface in the longitudinal direction is not
determined explicitly, but due to the factorization of the
transverse mass spectrum it can affect only the normalization. The
parameters connected with the expansion are the surface velocity
$\beta_{\perp}^{max}$ and $\rho_{max}$. The transverse velocity
profile has the following form

\begin{equation}
\beta_{\perp}(r)= \beta_{\perp}^{max}\left({ r \over
\rho_{max}}\right)^{\alpha}\;, \label{Betprof2}
\end{equation}
%%%%%%%%%%%%%%%%%%%%%%%%%%%%eq.20

\noindent with the choice $\alpha=1$. Therefore, the
implementation of values of $\beta_{\perp}^{max}$ obtained within
that model into the presented one is entirely \emph{ad hoc},
nevertheless it works surprisingly well. Of course, one directly
could apply the description of the transverse flow from
\cite{Schnedermann:1993ws} to calculate the transverse energy and
charged particle densities, but it is much more tempting and
elegant to work within one model. Additionally, there is one
technical problem connected with the treatment of resonance
decays. Here, the very convenient form of the invariant
distribution, eq.~(\ref{Cooper}), has been derived because the
normal vector is proportional to the four-velocity, $d\sigma_{\mu}
\propto u_{\mu}$. This is not the case for the hypersurface chosen
in \cite{Schnedermann:1993ws}, so the calculation of resonance
decay contributions would be much more complex (for the exact
formulae, see \cite{Broniowski:2002nf}). Note also that in the
view of \cite{Esumi:1997}, particle distributions depend very
weakly on the exact form of the velocity profile (\emph{i.e.} for
considered $\alpha=0.5,\;1,$ and $2$) in the model described in
\cite{Schnedermann:1993ws}. It can be checked that profile
(\ref{Betprof1}) lies in between two profiles of the form
(\ref{Betprof2}) with $\alpha=0.5,$ and $1$.
\begin{table}
\caption{\label{Table1} Comparison of the statistical model estimates of the
rapidity densities of charged particles with the experimental
values for the $5 \%$ most central collisions at
$\sqrt{s_{NN}}=130$ GeV at RHIC \protect\cite{Adcox:2001mf}.}
\begin{ruledtabular}
\begin{tabular}{ccc} \hline { }
& \multicolumn{2}{c}{} \\ Particles & \multicolumn{2}{c}{$
dN_{ch}/dy \vert_{y=0}$ }
\\
 & \multicolumn{2}{c}{} \\
\cline{2-3}  & Theory & Experiment
\\
 & \multicolumn{2}{c}{}
\\
\hline  & &
 \\
 $\pi^{+}+\pi^{-}$ & 548.6 & $546 \pm 54.5$
\\
 & &
\\
 $K^{+}+K^{-}$ & 84.6 & $87.2 \pm 12.0$
\\
 & & \\ $p+\bar{p}$
& 55.8 & $48.8 \pm 6.2$
\\
 & &
\\ \hline
\end{tabular}
\end{ruledtabular}
\end{table}

To put values of $\beta_{\perp}^{max}$ from
\cite{Braun-Munzinger:1994xr,Stachel:wh} into formulae of
sect.~\ref{density}, one should invert eq.~(\ref{Betmax}) to
obtain

\begin{equation}
{{\rho_{max}} \over {\tau} }=  { {\beta_{\perp}^{max}} \over
{\sqrt{1-(\beta_{\perp}^{max})^{2}}}}\;. \label{Rhmaxta}
\end{equation}
%%%%%%%%%%%%%%%%%%%%%%%%%%%%eq.21

\noindent It should be recalled here, that the value of $\tau$
itself is not necessary to calculate the transverse energy per
charged particle, since this parameter cancels in the ratio.

The final results of numerical estimates of
$dE_{T}/d\eta\vert_{mid}$ and $dN_{ch}/d\eta\vert_{mid}$ together
with the corresponding experimental data are listed in
TABLE\,\ref{Table2}. To make predictions for the AGS case it has
been assumed that the maximal transverse size $\rho_{max}$ equals
the average of radii of two colliding nuclei and the nucleus
radius has been expressed as $R_{A}=r_{0}A^{{1 \over
3}},\;r_{0}=1.12$ fm. Generally, the overall agreement is good.
For RHIC the $11\%-16\%$ underestimation of the charged particle
density has been received ($5\%-10\%$ with respect to the lowest
allowed values). But this result simply reflects the existing
inconsistency in measurements of the charged particle multiplicity
at RHIC. Namely, the sum of integrated charged hadron yields (see
TABLE\,\ref{Table1}), after converting to $dN_{ch}/d\eta$
\cite{Bazilevsky:2002fz}, is substantially less then the directly
measured $dN_{ch}/d\eta\vert_{mid}$. This is shown explicitly in
the last column of TABLE\,\ref{Table2}. For RHIC at
$\sqrt{s_{NN}}=130$ GeV, the sum is $8.7 \%$ smaller then the
total charged particle multiplicity. For RHIC at
$\sqrt{s_{NN}}=200$ GeV it is even worse, the sum is about $17 \%$
below the total $dN_{ch}/d\eta\vert_{mid}$. But both values of the
sum agree very well with the model predictions. Since the
geometric parameters were established from the fits to the same
$p_{T}$ spectra, the agreement had to be obtained. Also for AGS
the results agree qualitatively well with the data, in spite of
the roughness of the method applied for this case. The overall
error of evaluations of transverse energy and charged particle
densities is about $0.5\%$ and has two origins: (\textit{a})
omission of the most complex cascades; (\textit{b})
simplifications in numerical procedures for more involved
cascades. The velocity of the center of mass of two colliding
nuclei, $v_{c.m.s}$, equals: 0 for RHIC, 0.994 for SPS Pb-Pb
collisions at $158 \cdot A$ GeV, 0.918 for AGS Au-Au collisions at
$11 \cdot A$ GeV and 0.678 for AGS Si-Pb collisions at $14.6 \cdot
A$ GeV.
%%%%%%%%%%%%%%
\begin{table}
\caption{\label{Table2} Values of $dE_{T}/d\eta\vert_{mid}$ and
$dN_{ch}/d\eta\vert_{mid}$ calculated in the framework of the
statistical model with expansion. In the first column thermal and
geometric parameters are listed for the corresponding collisions.
In the third and last column experimental data for the most
central collisions are given.}
\begin{ruledtabular}
\begin{tabular}{c c c c c c} \hline Collision case &
\multicolumn{2}{c}{$dE_{T}/d\eta\vert_{mid}$
[GeV]} & & \multicolumn{2}{c}{$dN_{ch}/d\eta\vert_{mid}$}
\\
\cline{2-3}\cline{5-6} & Theory & Experiment & & Theory &
Experiment
\\
\hline Au-Au at RHIC at $\sqrt{s_{NN}}=200$ GeV: & & & & &
\\
 $T
= 165.6$ MeV, $\mu_{B} = 28.5$ MeV & 585 \footnote{For the
modified definition of $E_{T}$, \emph{i.e.} $E_{i}+m_{N}$ is taken
instead of $E_{i}$ for antibaryons, see
eq.~(\protect\ref{Etdef}).} & $597 \pm 34$
\protect\cite{Bazilevsky:2002fz} & & 589 & $699 \pm 46$
\protect\cite{Bazilevsky:2002fz}
\\
 $\rho_{max} = 7.15$ fm, $\tau
= 7.86$ fm ($\beta_{\perp}^{max} = 0.67$)
\protect\cite{Baran:2003nm} & & & & & $579 \pm 29$ \footnote{For
the charged particle multiplicity expressed as the sum of
integrated charged hadron yields.}
\\
 & & & & & \protect\cite{Adler:2003cb} \\  Au-Au at
RHIC at $\sqrt{s_{NN}}=130$ GeV: & & & & &
\\
 $T = 165$ MeV,
$\mu_{B} = 41$ MeV & 507 & $503 \pm 25$
\protect\cite{Adcox:2001ry} & & 555 & $622 \pm 41$
\protect\cite{Adcox:2000sp}
\\
 $\rho_{max} = 6.9$ fm, $\tau =
8.2$ fm ($\beta_{\perp}^{max} = 0.64$)
\protect\cite{Broniowski:2002nf} & & & & & $568 \pm 47\;^{b}$
\\
 & & & & & \protect\cite{Adcox:2001mf}
\\
 Pb-Pb
at SPS: & & & & & \\
 $T = 164$ MeV, $\mu_{B} = 234$ MeV & 447 &
$363 \pm 91$ \protect\cite{Aggarwal:2000bc} & & 476 &
$464_{-13}^{+20}$ \protect\cite{Aggarwal:2000bc}
\\ $\rho_{max} =
6.45$ fm, $\tau = 5.74$ fm ($\beta_{\perp}^{max} = 0.75$)
\protect\cite{Michalec:2001um,Broniowski:2002am} & & & & & \\  & &
& & & \\ Au-Au at AGS: & & & & & \\ $T = 130$ MeV, $\mu_{B} =
540$ MeV & 224 & $\approx 200$ \protect\cite{Barrette:pm} & & 271
& $\approx 270$ \protect\cite{Barrette:1994kr} \\
$\beta_{\perp}^{max} = 0.675$, $\rho_{max} = 6.52$ fm
\protect\cite{Braun-Munzinger:1994xr,Stachel:wh} & & & & & \\ & &
& & & \\  Si-Pb at AGS: & & & & & \\  $T = 120$ MeV, $\mu_{B} =
540$ MeV & 57 & $\approx 62$ \protect\cite{Barrette:1994kr} & & 91
& $\approx 115-120$ \\ $\beta_{\perp}^{max} = 0.54$, $\rho_{max} =
5.02$ fm \protect\cite{Braun-Munzinger:1994xr,Stachel:wh} & & & & &
\protect\cite{Barrette:1994kr} \\
\hline
\end{tabular}
\end{ruledtabular}
\end{table}

Values of the ratio $dE_{T}/d\eta\vert_{mid} /
dN_{ch}/d\eta\vert_{mid}$ can be also given. They are collected in
TABLE\,\ref{Table3}, together with the corresponding data.
Generally, the overall overestimation of the order of $15\%$ has
been obtained. In the RHIC case this is the result of the
underestimation of $dN_{ch}/d\eta\vert_{mid}$, which has been
explained earlier. But when in the denominator of the experimental
ratio, $dN_{ch}/d\eta\vert_{mid}$ from the summing up of
integrated hadron yields is put, the theoretical predictions agree
very well with the data. Note that the similar inconsistency in
charged particle measurements could have also been the origin of
the discrepancy between model and experimental values of
$dN_{ch}/d\eta\vert_{mid}$ seen in the AGS Si-Pb case. For SPS,
the result agrees with the experimental value within errors. The
overall error of model evaluations of the ratio is less than $1
\%$.
\begin{table}
\caption{\label{Table3} Values of the ratio
$dE_{T}/d\eta\vert_{mid}/ dN_{ch}/d\eta\vert_{mid}$ calculated in
the framework of the statistical model with expansion. In the last
column experimental data for the most central collisions are
given.}
\begin{ruledtabular}
\begin{tabular}{c c c} \hline { }
& \multicolumn{2}{c}{} \\ Collision case & \multicolumn{2}{c}{
$dE_{T}/d\eta\vert_{mid}/ dN_{ch}/d\eta\vert_{mid}$ [GeV]} \\  &
\multicolumn{2}{c}{} \\ \cline{2-3} & Theory & Experiment \\  &
\multicolumn{2}{c}{} \\ \hline & & \\ Au-Au at RHIC at
$\sqrt{s_{NN}}=200$ GeV & 0.99 \footnote{For the modified
definition of $E_{T}$, \emph{i.e.} $E_{i}+m_{N}$ is taken instead
of $E_{i}$ for antibaryons, see eq.~(\protect\ref{Etdef}).} &
$0.87 \pm 0.06$ \protect\cite{Bazilevsky:2002fz} \\ & & $1.03 \pm
0.08$ \footnote{Author calculations with the use of experimental
values given in TABLE\,\protect\ref{Table2} and the denominator
expressed as the sum of integrated charged hadron yields.} \\
Au-Au at RHIC at $\sqrt{s_{NN}}=130$ GeV & 0.91 & $0.81 \pm 0.06$
\protect\cite{Adcox:2001ry} \\ & & $0.89 \pm 0.09\;^{b}$ \\ Pb-Pb
at SPS
& 0.94 & $0.78 \pm 0.21$ \protect\cite{Aggarwal:2000bc} \\ & & \\
Au-Au
at AGS & 0.83 & $0.72 \pm 0.08$ \protect\cite{Barrette:1994kr} \\
& & \\ Si-Pb at AGS & 0.63 & 0.52-0.54
\protect\cite{Barrette:1994kr} \\ & & \\ \hline
\end{tabular}
\end{ruledtabular}
\end{table}
\begin{figure}
\includegraphics{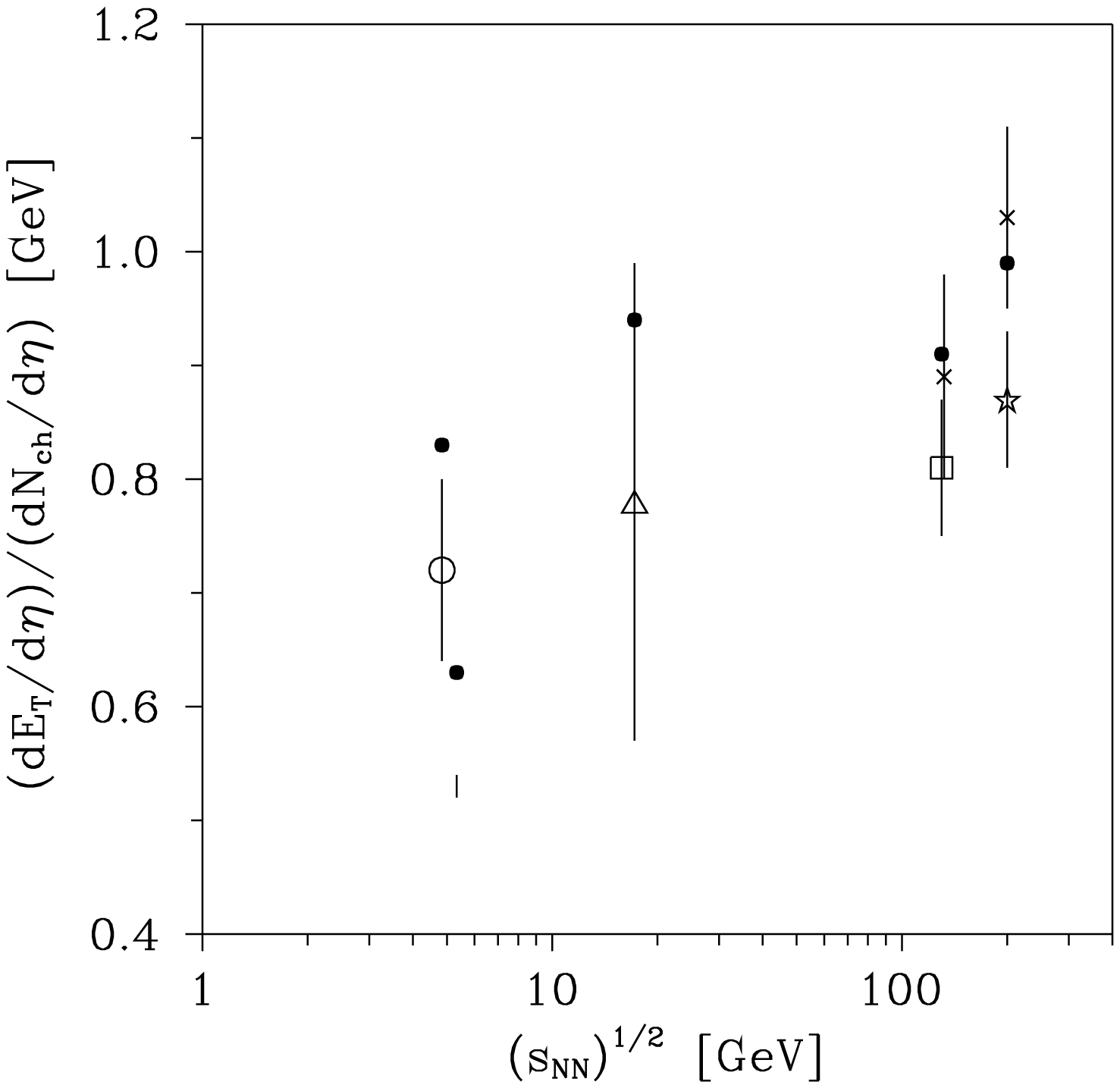}
\caption{\label{Fig.1.} Values of the transverse energy per
charged particle at midrapidity for the most central collisions.
Black dots denote evaluations of the ratio in the framework of the
present model (the second column of TABLE\,\protect\ref{Table3}).
Also data points for AGS \protect\cite{Barrette:1994kr} (a circle
for Au-Au and a vertical bar for Si-Pb), SPS
\protect\cite{Aggarwal:2000bc} (triangle), RHIC at
$\sqrt{s_{NN}}=130$ GeV \protect\cite{Adcox:2001ry} (square) and
RHIC at $\sqrt{s_{NN}}=200$ GeV \protect\cite{Bazilevsky:2002fz}
(star) are depicted. For RHIC, points with the sum of integrated
charged hadron yields substituted for the denominator are also
depicted (crosses).}
\end{figure}
These results have been also depicted together with the data in
fig.\,\ref{Fig.1.}. One can see that the relative positions of
theoretical points agree very well with the data, they are shifted
up only and this is the effect of the overestimation discussed
earlier.

It should be stressed that $dE_{T}/d\eta\vert_{mid}/
dN_{ch}/d\eta\vert_{mid}$ depends substantially on the value of
$\beta_{\perp}^{max}$. It can be seen from TABLE\,\ref{Table4},
where the transverse energy per charged particle estimates have
been listed for a few values of $\beta_{\perp}^{max}$ for Au-Au
collisions at AGS. Having compared with the experimental data (see
TABLE\,\ref{Table3}), one can notice that this model yields the
value of $\beta_{\perp}^{max}$ which is slightly lower than the
value obtained within the model described in
\cite{Schnedermann:1993ws}.
\begin{table}
\caption{\label{Table4} Dependence of the transverse energy per charged particle
on the maximum transverse-flow parameter $\beta_{\perp}^{max}$ for
Au-Au collisions at AGS.}
\begin{ruledtabular}
\begin{tabular}{c c} \hline & \\
$\beta_{\perp}^{max}$ & $dE_{T}/d\eta\vert_{mid}/
dN_{ch}/d\eta\vert_{mid}$ [GeV] \\  \hline & \\  0.4 & 0.61 \\
& \\  0.57 & 0.72 \\ & \\  0.675 & 0.83 \\  & \\  0.8 & 1.07 \\  &
\\  \hline
\end{tabular}
\end{ruledtabular}
\end{table}

\section {Comparison with a static case}
\label{Staticc}

It would be very interesting to check how expansion influences the
transverse energy per charged particle. The expansion produces
additional energy so this process should increase the energy of a
particle emitted from a thermal source. The preliminary analysis
of a static case was done in \cite{Prorok:2002ta}. But to compare
with the present results up-to-date calculations should be
performed. In \cite{Prorok:2002ta} a gas with only $40$ species
(including antiparticles) was examined and feeding charged
particles from weak decays of neutral resonances was excluded.
Thus, to extract the expansion contribution to
$dE_{T}/d\eta\vert_{mid} / dN_{ch}/d\eta\vert_{mid}$ one has to
apply the general scheme of sect.\,\ref{density} again, but with
the proper replacement of the invariant particle distribution.

For a static gas (\emph{static} in the c.m.s, of course),  the
invariant distribution of the measured particles of species $i$
has the form \cite{Prorok:2002ta}

\begin{equation}
{ {dN_{i}} \over {d^{2}p_{T}\;dy} }=V^{*}\; E_{i}^{*}
\;f_{i}(E_{i}^{*})\;, \label{Coosta}
\end{equation}
%%%%%%%%%%%%%%%%%%%%%%%%%%%%eq.22

\noindent where $E_{i}^{*}$ is the c.m.s energy of the \emph{i}th
particle and $V^{*}$ denotes the c.m.s volume of the gas at the
freeze-out. Thus, at midrapidity, one has:

\begin{equation}
{ {dN_{i}} \over {d^{2}p_{T}\;dy} }\;\Big\vert_{mid}=V^{*}\; m_{T}
\;f_{i}(m_{T})\;. \label{Coostamid}
\end{equation}
%%%%%%%%%%%%%%%%%%%%%%%%%%%%eq.23

Now the general formulae of eqs.~(\ref{Partdenmid})-(\ref{Etall})
can be applied, but with $dN_{i}/d^{2}p_{T}dy$ given by
eq.~(\ref{Coostamid}) instead of eq.~(\ref{Cooper2}). The results
of numerical evaluations of $dE_{T}/d\eta\vert_{mid} /
dN_{ch}/d\eta\vert_{mid}$ for the static gas are collected in
TABLE\,\ref{Table5}. In this case only two (thermal) parameters
are needed and they are the same as in TABLE\,\ref{Table2}. Having
compared with TABLE\,\ref{Table3}, one can see that expansion is
responsible for the following increases of the transverse energy
per charged particle: $11\%$ for RHIC, $32\%$ for SPS, $34\%$ for
AGS Au-Au collisions and $19\%$ for AGS Si-Pb collisions. This can
be explained reasonably. The transverse energy per charged
particle has two contributions: the first thermal and the second
originated from expansion.
\begin{table}
\caption{\label{Table5} Values of the ratio
$dE_{T}/d\eta\vert_{mid}/ dN_{ch}/d\eta\vert_{mid}$ calculated for
the static gas. In the last column experimental data for the most
central collisions are given.}
\begin{ruledtabular}
\begin{tabular}{c c c} \hline { }
& \multicolumn{2}{c}{} \\  Collision case &
\multicolumn{2}{c}{$dE_{T}/d\eta\vert_{mid}/
dN_{ch}/d\eta\vert_{mid}$ [GeV]} \\  & \multicolumn{2}{c}{}
 \\ \cline{2-3} & Theory & Experiment \\  & \multicolumn{2}{c}{}
\\  \hline & &
 \\  Au-Au at RHIC at $\sqrt{s_{NN}}=200$ GeV & 0.89
\footnote{For the modified definition of $E_{T}$, \emph{i.e.}
$E_{i}+m_{N}$ is taken instead of $E_{i}$ for antibaryons, see
eq.~(\protect\ref{Etdef}).} &
 $0.87 \pm 0.06$ \protect\cite{Bazilevsky:2002fz}
\\  & & \\  Au-Au at RHIC at $\sqrt{s_{NN}}=130$ GeV & 0.82 &
 $0.81 \pm 0.06$ \protect\cite{Adcox:2001ry}
\\ & & \\ Pb-Pb at SPS & 0.71 & $0.78 \pm 0.21$
\protect\cite{Aggarwal:2000bc} \\  & & \\ Au-Au at AGS & 0.62 &
$0.72 \pm 0.08$ \protect\cite{Barrette:1994kr} \\  & & \\  Si-Pb
at AGS & 0.53 & 0.52-0.54 \protect\cite{Barrette:1994kr} \\  & &
\\
\hline
\end{tabular}
\end{ruledtabular}
\end{table}
The first is governed mainly by the temperature and the second by
the maximum transverse-flow parameter $\beta_{\perp}^{max}$. For a
given temperature, the increase of $\beta_{\perp}^{max}$ should
cause the weighting of the expansion contribution. But for a
constant value of $\beta_{\perp}^{max}$, the strengthening of this
contribution can be maintained by the lowering of the temperature.
This is why for almost the same $\beta_{\perp}^{max}$ (see the
first column of TABLE\,\ref{Table2}) the relative growth of the
transverse energy per charged particle, after switching the
expansion on, is much greater for AGS Au-Au collisions than for
RHIC ones. On the other hand, for comparable temperatures, the
expansion contributes to the transverse energy per charged
particle much stronger for SPS than for RHIC (the former has
substantially greater $\beta_{\perp}^{max}$).

\section {Conclusions}
\label{Conclud}

The expanding thermal hadron gas model has been used to reproduce
transverse energy and charged particle multiplicity pseudorapidity
densities and their ratio measured at AGS, SPS and RHIC. The
importance of the present analysis originates from the fact that
the transverse energy and the charged particle multiplicity are
\emph{independent observables}, so they can be used as new tools
to verify the consistency of predictions of a statistical model
for all colliders simultaneously. The predictions have been made
at the previous estimates of thermal and geometric freeze-out
parameters obtained from analyses of measured particle ratios and
$p_{T}$ spectra at AGS \cite{Braun-Munzinger:1994xr,Stachel:wh},
SPS \cite{Michalec:2001um,Broniowski:2002am} and RHIC
\cite{Baran:2003nm,Broniowski:2002nf}. The overall good agreement,
not only of the ratio but also absolute values of
$dE_{T}/d\eta\mid_{mid}$ and $dN_{ch}/d\eta\vert_{mid}$, with the
data has been achieved. And the observed discrepancies can be
explained reasonably. This strongly supports the idea that the
thermal expanding source is responsible for the soft part of the
particle production in heavy-ion collisions. Moreover, the
description of various observables is consistent within one
statistical model.

In fact, there are additional arguments which make the above
statement even more valuable. In principle, one could think at
first glance that this analysis is nothing more like a kind of an
internal consistency check of various measurements. And such a
check could be done even in an model-independent way simply by
integrating spectra of stable particles (the first time with the
expression for transverse energy to obtain
$dE_{T}/d\eta\mid_{mid}$ and the second time without, to receive
$dN_{ch}/d\eta\vert_{mid}$) and then adding them all. But there
are two reasons that this can not be done without any external
input. First, transverse momentum spectra are measured in
\emph{limited ranges}, so very important low-$p_{T}$ regions are
not covered by the data. For instance at RHIC, the first point for
pions is at $p_{T}= 0.25$ GeV/c, for kaons at $p_{T}= 0.45$ GeV/c
and for protons and antiprotons at $p_{T}= 0.65$ GeV/c
\cite{Adcox:2001mf,Adler:2003cb}. There are also upper limits, but
contributions from ranges above them are suppressed strongly in
comparison with the low-$p_{T}$ regions. Therefore, to obtain
integrated yields some extrapolations below and above the measured
ranges are used. Usually two functions are used for each species
and the contributed value is the average of their integrals. In
fact these extrapolations are only analytical fits without any
physical reasoning, but, for instance, contributions from regions
covered by them account for $30 \%$ of the yield for pions, $40
\%$ for kaons and $25 \%$ for protons and antiprotons for RHIC at
$\sqrt{s_{NN}}=130$ GeV \cite{Adcox:2001mf}. On the other hand, a
calorimeter acts very effectively for these species in the
low-$p_{T}$ range, namely pions with $p_{T} \leq 0.35$ GeV/c,
kaons with $p_{T} \leq 0.64$ GeV/c and protons and antiprotons
with $p_{T} \leq 0.94$ GeV/c deposit all their kinetic energy
\cite{Adcox:2001ry}. Since the very accurate predictions for the
transverse energy density at midrapidity have been obtained (see
TABLE\,\ref{Table2}), the present analysis can be understood as an
undirect proof that in these unmeasurable $p_{T}$ regions spectra
are also explicable by means of the thermal source with flow and
decays.

Second, it is impossible to check the consistency of the
transverse energy data because not all stable hadron spectra are
measured at midrapidity for each collision case. This mainly
concerns neutrons and $K_{L}^{0}$. The lacking contribution from
hadron decay photons could be approximated to some extent with the
use of $\pi^{0}$ and $\eta$ spectra, but they are also limited in
ranges. And again, the very good agreement of model estimates of
the transverse energy density at midrapidity with the data can be
interpreted as the strong argument that the production of neutral
stable particles can be described in terms of the expanding
thermal source with superimposed decays.

And last but not least, in opposite to the transverse energy,
there is some inconsistency (of the order of $10 \%$) of the
independent measurements of charged particle multiplicities with
the corresponding sums of integrated charged particle yields at
RHIC (see sect.~\ref{Finl}). However, only for the case of
$\sqrt{s_{NN}}=200$ GeV the substantial gap (of the order of $6
\%$ with respect to the direct measurement) between error bars of
these two differently obtained values of
$dN_{ch}/d\eta\vert_{mid}$ exists. For the case of
$\sqrt{s_{NN}}=130$ GeV the error bars overlap almost one half of
each other. But since the data at $\sqrt{s_{NN}}=200$ GeV are
still preliminary, it is difficult to judge whether this
inconsistency has the physical or experimental (an additional
systematic error?) reason.

The role of expansion is substantial. It produces about
$10\%-30\%$ of the transverse energy per charged particle. But, as
can be seen from TABLE\,\ref{Table5}, the expansion is not
necessary to explain the experimental data for
$dE_{T}/d\eta\mid_{mid} / dN_{ch}/d\eta\vert_{mid}$. The results
suggest that the most of the transverse energy per charged
particle is produced by the thermal movement. For sure, the
expansion is necessary to explain the absolute values of
$dE_{T}/d\eta\mid_{mid}$ and $dN_{ch}/d\eta\vert_{mid}$. To obtain
these one needs a volume of a place of "action" and the most
adequate way to do it is to parameterize the evolution of the
system in space and time, that is to put the expansion in.

As the last comment, it should be stressed that the results of the
present paper have been obtained within the model where the
chemical freeze-out happens simultaneously with the thermal one.
However, so far the most extensively studied scenario is that
where the thermal freeze-out occurs later then the chemical
freeze-out (for a review, see \cite{Heinz:1999kb} and references
therein). This problem has not been addressed here. But one should
notice that the distinction between these two freeze-outs means
the introduction of the next parameter (the fifth here) into the
model. Of course, an extra parameter in a phenomenological model
always causes (or at least should cause) better agreement with the
data. At the present level of investigations both spectra (refs.~
\cite{Broniowski:2001we,Broniowski:2001uk,Broniowski:2002am}) and
global observables $dE_{T}/d\eta\mid_{mid}$ and
$dN_{ch}/d\eta\vert_{mid}$ (this analysis) are predicted
accurately with the assumption of the one freeze-out. However,
more detailed studies should be performed to check whether the
transverse energy per charged particle measurement could help
somehow in distinction or not between these two freeze-outs and
this will be the subject of further investigations.

\begin{acknowledgments}
The author gratefully acknowledges very stimulating discussions
with Wojciech Broniowski and Wojciech Florkowski. He also thanks
Ludwik Turko for careful reading of the manuscript and Anna
Jadczyk for help in preparing the \LaTeXe file. This work was
supported in part by the Polish Committee for Scientific Research
under Contract No. KBN 2 P03B 069 25.
\end{acknowledgments}

\appendix*
\section{}

The derivation of the momentum distribution of a product of a
two-body decay $M \longrightarrow m_{1} + m_{2}$ can be found in
\cite{Hagedorn:1973} or \cite{Florkowski:2001fp}. For an $n$-body
decay $M \longrightarrow m_{1} + m_{2}+...+m_{n}$, the momentum
distribution of the product (labeled 1) can be written (following
the method presented in \cite{Florkowski:2001fp}) as:

\begin{eqnarray}
f^{(n)}_{1}(\vert\vec{q}\vert,M,m_{1},m_{2},...,m_{n}) &=&
B{{2s_{M}+1} \over {2s_{1}+1}}{1 \over
{N^{(n)}(M;m_{1},m_{2},...,m_{n})}} \int d^{3}\vec{k}
f_{M}(\vert\vec{k}\vert) \cr \cr &&\times \int
\left(\prod_{i=1}^{n} { {d^{3}\vec{p}_{i}} \over {E_{i}}}\right)\;
\delta(M-\sum_{i=1}^{n}E_{i})
\delta^{(3)}(\sum_{i=1}^{n}\vec{p}_{i}) \cr \cr &&\times
\delta^{(3)}(\hat{L}_{\vec{k}}\vec{p}_{1}-\vec{q})\;,
\label{distnb}
\end{eqnarray}
%%%%%%%%%%%%%%%%%%%%%%%%%%%%eq.A1

\noindent where

\begin{equation}
\hat{L}_{\vec{k}}\vec{p}_{1}=\vec{p}_{1}+\left\{(\gamma_{k}-1) {
{\vec{p}_{1} \cdot \vec{k}} \over {k^{2}}} +  { {E_{1}} \over
M}\right\}\vec{k}\;,\label{Lorentz}
\end{equation}
%%%%%%%%%%%%%%%%%%%%%%%%%%%%eq.A2

\begin{equation}
\gamma_{k} = { {E_{M}} \over M}\;,\;\;\;E_{M}=
\sqrt{M^{2}+\vec{k}^{2}}\;,\;\;\;E_{i}=
\sqrt{m^{2}_{i}+\vec{p}^{2}_{i}}\;,\label{gama}
\end{equation}
%%%%%%%%%%%%%%%%%%%%%%%%%%%%eq.A3

\noindent and $s_{M}$ ($s_{1}$) is the spin of the resonance (the
product), $B$ is the branching ratio, $f_{M}(\vert\vec{k}\vert)$
denotes the momentum distribution of the decaying resonance and
$N^{(n)}$ is the corresponding phase-space integral:

\begin{equation}
N^{(n)}(M;m_{1},m_{2},...,m_{n})= \int \left(\prod_{i=1}^{n} {
{d^{3}\vec{p}_{i}} \over {E_{i}}}\right)\;
\delta(M-\sum_{i=1}^{n}E_{i})
\delta^{(3)}(\sum_{i=1}^{n}\vec{p}_{i})\;. \label{volnb}
\end{equation}
%%%%%%%%%%%%%%%%%%%%%%%%%%%%eq.A4

\noindent The invariant amplitude for the decay, ${\cal M}$, is
assumed to be a constant here, so $\vert {\cal M}\vert^{2}$
cancels during normalization.

With the use of the well-known technique of splitting up the
phase-space integral into a convolution integral over two
phase-space integrals (here, the first responsible for the
$2$-body decay and the second representing the $(n-1)$-body decay)
\cite{Hagedorn:1973}, the following recursive formulae for the
$n$-body decay can be derived:
%\newpage
\begin{eqnarray}
f^{(n)}_{1}(\vert\vec{q}\vert,M,m_{1},m_{2},...,m_{n}) &=&
B{{2s_{M}+1} \over {2s_{1}+1}} {{2\pi} \over
{N^{(n)}(M;m_{1},m_{2},...,m_{n})}} {1 \over {qE_{1}(q)}} \cr \cr
&&\times \int\limits_{m_{2}+...+m_{n}}^{M-m_{1}} dm\;m
N^{(n-1)}(m;m_{2},...,m_{n}) \cr \cr &&\times
\int\limits_{k_{-}(q;M,m_{1},m)}^{k_{+}(q;M,m_{1},m)}
dk\;k\;f_{M}(k)\;, \label{dist3bb}
\end{eqnarray}
%%%%%%%%%%%%%%%%%%%%%%%%%%%%eq.A5

\begin{equation}
N^{(n)}(M;m_{1},m_{2},...,m_{n}) = {{4\pi} \over M}
\int\limits_{m_{2}+...+m_{n}}^{M-m_{1}} dm\;m
\;N^{(n-1)}(m;m_{2},...,m_{n}) \;p(M;m_{1},m)\;, \label{dist4bb}
\end{equation}
%%%%%%%%%%%%%%%%%%%%%%%%%%%%eq.A6

\noindent where

\begin{equation}
k_{\pm}(q;M,m_{1},m_{2})= {M \over {m_{1}^{2}}}\vert
p(M;m_{1},m_{2})E_{1}(q)\pm qE(M;m_{1},m_{2})\vert \;,
\label{limtk2b}
\end{equation}
%%%%%%%%%%%%%%%%%%%%%%%%%%%%eq.A7

\noindent for $m_{1} \neq 0$, whereas

\begin{equation}
\cases{k_{+}(q;M,m_{1},m_{2})= +\infty, \cr \cr
k_{-}(q;M,m_{1},m_{2})= { {\vert {1 \over 4}
(M^{2}-m_{2}^{2})^{2}-M^{2}q^{2}\vert} \over {(M^{2}-m_{2}^{2})q}
} } \; \label{limtk2b0}
\end{equation}
%%%%%%%%%%%%%%%%%%%%%%%%%%%%eq.A8

\noindent for $m_{1} = 0$, and

\begin{equation}
p(M;m_{1},m_{2})= {M \over {4\pi}} N^{(2)}(M;m_{1},m_{2})= {
{\sqrt{[M^{2}-(m_{1}+m_{2})^{2}][M^{2}-(m_{1}-m_{2})^{2}]}} \over
{2M}} \;,\label{pmax2b}
\end{equation}
%%%%%%%%%%%%%%%%%%%%%%%%%%%%eq.A9

\begin{equation}
E(M;m_{1},m_{2})= { {M^{2}-m_{2}^{2}+m_{1}^{2}} \over
{2M}}\;.\label{Emax2b}
\end{equation}
%%%%%%%%%%%%%%%%%%%%%%%%%%%%eq.A10


\begin{thebibliography}{99}
\bibitem{Prorok:2002ta}
D.~Prorok,
 %``Analysis of the freeze-out parameters for RHIC, SPS and AGS based on
%(d(E(T))/d(eta))/(d(N(ch))/d(eta)) ratio measurement,''
Acta Phys.\ Polon.\ B {\bf 34}, 4219 (2003).
%%CITATION = HEP-PH 0212103;%%

\bibitem{Braun-Munzinger:1994xr}
P.~Braun-Munzinger, J.~Stachel, J.~P.~Wessels and N.~Xu,
%``Thermal equilibration and expansion in nucleus-nucleus collisions at the AGS,''
Phys.\ Lett.\ B {\bf 344}, 43 (1995).
%%CITATION = NUCL-TH 9410026;%%

\bibitem{Braun-Munzinger:1995bp}
P.~Braun-Munzinger, J.~Stachel, J.~P.~Wessels and N.~Xu,
 %``Thermal and hadrochemical equilibration in nucleus-nucleus collisions at the
%SPS,''
Phys.\ Lett.\ B {\bf 365}, 1 (1996).
%%CITATION = NUCL-TH 9508020;%%

\bibitem{Cleymans:1996cd}
J.~Cleymans, D.~Elliott, H.~Satz and R.~L.~Thews,
%``Thermal hadron production in Si-Au collisions,''
Z.\ Phys.\ C {\bf 74}, 319 (1997).
%%CITATION = NUCL-TH 9603004;%%

\bibitem{Stachel:wh}
J.~Stachel,
%``Tests Of Thermalization In Relativistic Nucleus Nucleus Collisions,''
Nucl.\ Phys.\ A {\bf 610}, 509C (1996).
%%CITATION = NUPHA,A610,509C;%%

\bibitem{Braun-Munzinger:1999qy}
P.~Braun-Munzinger, I.~Heppe and J.~Stachel,
%``Chemical equilibration in Pb + Pb collisions at the SPS,''
Phys.\ Lett.\ B {\bf 465}, 15 (1999).
%%CITATION = NUCL-TH 9903010;%%

\bibitem{Becattini:2000jw}
F.~Becattini, J.~Cleymans, A.~Keranen, E.~Suhonen and K.~Redlich,
 %``Features of particle multiplicities and strangeness production in  central
%heavy ion collisions between 1.7-A-GeV/c and 158-A-GeV/c,''
Phys.\ Rev.\ C {\bf 64}, 024901 (2001).
%%CITATION = HEP-PH 0002267;%%

\bibitem{Braun-Munzinger:2001ip}
P.~Braun-Munzinger, D.~Magestro, K.~Redlich and J.~Stachel,
%``Hadron production in Au Au collisions at RHIC,''
Phys.\ Lett.\ B {\bf 518}, 41 (2001).
%%CITATION = HEP-PH 0105229;%%

\bibitem{Florkowski:2001fp}
W.~Florkowski, W.~Broniowski and M.~Michalec,
%``Thermal analysis of particle ratios and p(T) spectra at RHIC,''
Acta Phys.\ Polon.\ B {\bf 33}, 761 (2002).
%%CITATION = NUCL-TH 0106009;%%

\bibitem{Broniowski:2001we}
W.~Broniowski and W.~Florkowski,
%``Explanation of the RHIC p(T)-spectra in a thermal model with expansion,''
Phys.\ Rev.\ Lett.\  {\bf 87}, 272302 (2001).
%%CITATION = NUCL-TH 0106050;%%

\bibitem{Broniowski:2001uk}
W.~Broniowski and W.~Florkowski,
%``Description of strange particle production in Au+Au collisions of sNN =130 GeV in a single-freeze-out model,''
Phys.\ Rev.\ C {\bf 65}, 064905 (2002).
%%CITATION = NUCL-TH 0112043;%%

\bibitem{Baran:2003nm}
A.~Baran, W.~Broniowski and W.~Florkowski,
 %``Description of the particle ratios and transverse-momentum spectra for
%various centralities at RHIC in a single-freeze-out model,''
Acta Phys.\ Polon.\ B {\bf 35}, 779 (2004).
%%CITATION = NUCL-TH 0305075;%%

\bibitem{Broniowski:2002nf}
W.~Broniowski, A.~Baran and W.~Florkowski,
%``Thermal approach to RHIC,''
Acta Phys.\ Polon.\ B {\bf 33}, 4235 (2002).
%%CITATION = HEP-PH 0209286;%%

\bibitem{Michalec:2001um}
M.~Michalec, Ph.D. thesis,
%``Thermal description of particle production in ultra-relativistic  heavy-ion collisions,''
arXiv:nucl-th/0112044.
%%CITATION = NUCL-TH 0112044;%%

\bibitem{Broniowski:2002am}
W.~Broniowski and W.~Florkowski,
%``How much is RHIC different from SPS? Comparison of the p(T)-spectra,''
Acta Phys.\ Polon.\ B {\bf 33}, 1935 (2002).
%%CITATION = NUCL-TH 0204025;%%

\bibitem{Torrieri:2004zz}
G.~Torrieri, W.~Broniowski, W.~Florkowski, J.~Letessier and
J.~Rafelski,
%``SHARE: Statistical hadronization with resonances,''
arXiv:nucl-th/0404083.
%%CITATION = NUCL-TH 0404083;%%

\bibitem{Wheaton:2004qb}
S.~Wheaton and J.~Cleymans,
%``THERMUS: A thermal model package for ROOT,''
arXiv:hep-ph/0407174.
%%CITATION = HEP-PH 0407174;%%

\bibitem{Back:2000gw}
B.~B.~Back {\it et al.}  [PHOBOS Collaboration],
%``Charged particle multiplicity near mid-rapidity in central Au + Au
%collisions at s**(1/2) = 56-A/GeV and 130-A/GeV,''
Phys.\ Rev.\ Lett.\  {\bf 85}, 3100 (2000).
%%CITATION = HEP-EX 0007036;%%

\bibitem{Adcox:2001ry}
K.~Adcox {\it et al.}  [PHENIX Collaboration],
%``Measurement of the mid-rapidity transverse energy distribution from  s(N N)**(1/2) = 130-GeV Au + Au collisions at RHIC,''
Phys.\ Rev.\ Lett.\  {\bf 87}, 052301 (2001).
%%CITATION = NUCL-EX 0104015;%%

\bibitem{Bazilevsky:2002fz}
A.~Bazilevsky  [PHENIX Collaboration],
 %``Charged particle multiplicity and transverse energy measurements in Au Au
%collisions in PHENIX at RHIC,''
Nucl.\ Phys.\ A {\bf 715}, 486 (2003).
%%CITATION = NUCL-EX 0209025;%%

\bibitem{Hagiwara:fs}
K.~Hagiwara {\it et al.}  [Particle Data Group Collaboration],
%``Review Of Particle Physics,''
Phys.\ Rev.\ D {\bf 66}, 010001 (2002).
%%CITATION = PHRVA,D66,010001;%%

\bibitem{Cooper:1974mv}
F.~Cooper and G.~Frye,
%``Single Particle Distribution In The Hydrodynamic And Statistical Thermodynamic Models Of Multiparticle Production,''
Phys.\ Rev.\ D {\bf 10}, 186 (1974).
%%CITATION = PHRVA,D10,186;%%

\bibitem{Adcox:2000sp}
K.~Adcox {\it et al.}  [PHENIX Collaboration],
 %``Centrality dependence of charged particle multiplicity in Au Au  collisions
%at s(N N)**(1/2) = 130-GeV,''
Phys.\ Rev.\ Lett.\  {\bf 86}, 3500 (2001).
%%CITATION = NUCL-EX 0012008;%%

\bibitem{Adcox:2001mf}
K.~Adcox {\it et al.}  [PHENIX Collaboration],
 %``Centrality dependence of pi+-, K+-, p and anti-p production from
%s(NN)**(1/2) = 130-GeV Au + Au collisions at RHIC,''
Phys.\ Rev.\ Lett.\  {\bf 88}, 242301 (2002).
%%CITATION = NUCL-EX 0112006;%%

\bibitem{Schnedermann:1993ws}
E.~Schnedermann, J.~Sollfrank and U.~Heinz,
%``Thermal phenomenology of hadrons from 200-A/GeV S+S collisions,''
Phys.\ Rev.\ C {\bf 48}, 2462 (1993).
%%CITATION = NUCL-TH 9307020;%%

\bibitem{Esumi:1997}
S.~Esumi, S.~Chapman, H.~van Hecke and N.~Xu,
%``Transverse flow at ultrarelativistic energies,''
Phys.\ Rev.\ C {\bf 55}, R2163 (1997).

\bibitem{Aggarwal:2000bc}
M.~M.~Aggarwal {\it et al.}  [WA98 Collaboration],
%``Scaling of particle and transverse energy production in 208-Pb + 208-Pb  collisions at 158-A-GeV,''
Eur.\ Phys.\ J.\ C {\bf 18}, 651 (2001).
%%CITATION = NUCL-EX 0008004;%%

\bibitem{Barrette:pm}
J.~Barrette {\it et al.}  [E814/E877 Collaboration],
 %``Measurement Of Transverse Energy Production With Si And Au Beams At
%Relativistic Energy: Towards Hot And Dense Hadronic Matter,''
Phys.\ Rev.\ Lett.\  {\bf 70}, 2996 (1993).
%%CITATION = PRLTA,70,2996;%%

\bibitem{Barrette:1994kr}
J.~Barrette {\it et al.}  [E877 Collaboration],
%``Charged particle pseudorapidity distributions in Au + Al, Cu, Au, and U collisions at 10.8-A/GeV/c,''
Phys.\ Rev.\ C {\bf 51}, 3309 (1995).
%%CITATION = NUCL-EX 9412003;%%

\bibitem{Adler:2003cb}
S.~S.~Adler {\it et al.}  [PHENIX Collaboration],
%``Identified charged particle spectra and yields in Au + Au collisions at
%s(NN)**(1/2) = 200-GeV,''
Phys.\ Rev.\ C {\bf 69}, 034909 (2004).
%%CITATION = NUCL-EX 0307022;%%

\bibitem{Heinz:1999kb}
U.~W.~Heinz,
%``Primordial hadrosynthesis in the little bang,''
Nucl.\ Phys.\ A {\bf 661}, 140 (1999).
%%CITATION = NUCL-TH 9907060;%%

\bibitem{Hagedorn:1973}
R.~Hagedorn, {\it Relativistic kinematics}, (W.~A.~Benjamin, Inc.,
Advanced Book Program, Reading, Massachusetts, Third printing,
1973).
\end{thebibliography}
\end{document}